\documentclass[aps, pra, reprint, superscriptaddress]{revtex4-2}

\pdfoutput=1

\usepackage{amsmath}    
\usepackage{graphicx}   
\usepackage{subfigure}  
\usepackage{hyperref}   
\usepackage{xcolor}
\usepackage{multirow}	

\newcommand{\spinion}{Spin-Ion Technologies, 10 Boulevard Thomas Gobert, 91120 Palaiseau, France}
\newcommand{\parissaclay}{Universit\'e Paris-Saclay, 3 rue Juliot Curie, 91190 Gif-sur-Yvette, France}
\newcommand{\nanosciences}{Centre de Nanosciences et de Nanotechnologies, CNRS, Universit\'e Paris-Saclay, 10 Boulevard Thomas Gobert, 91120 Palaiseau, France}
\newcommand{\univparis}{Laboratoire des Sciences des Proc\'ed\'es et des Mat\'eriaux, CNRS-UPR 3407, Universit\'e Sorbonne Paris Nord, 93430 Villetaneuse, France}
\newcommand{\CNR}{CNR-IMM, Unit of Agrate Brianza, via C. Olivetti 2, 20864 Agrate Brianza, Italy}
\newcommand{\LPS}{Laboratoire de Physique des Solides, Universit\'e Paris-Saclay, CNRS, 91405 Orsay, France}
\newcommand{\USAL}{Departamento de F\'isica Aplicada, Universidad de Salamanca, Plaza de la Merced s/n, 37008 Salamanca, Spain}

\begin{document}

\title{Revealing nanoscale disorder in W/CoFeB/MgO ultra-thin films using domain wall motion}

\author{Johannes W. \surname{van der Jagt}}
\email{gyan.vdjagt@spin-ion.com}
\affiliation{\spinion}
\affiliation{\parissaclay}
\author{Vincent \surname{Jeudy}}
\author{Andr\'e \surname{Thiaville}}
\affiliation{\LPS}
\author{Mamour \surname{Sall}}
\affiliation{\spinion}
\author{Nicolas \surname{Vernier}}
\author{Liza \surname{Herrera Diez}}
\affiliation{\nanosciences}
\author{Mohamed \surname{Belmeguenai}}
\author{Yves \surname{Roussign\'e}}
\author{Salim M. \surname{Ch\'erif}}
\affiliation{\univparis}
\author{Mouad \surname{Fattouhi}}
\author{Luis \surname{Lopez-Diaz}}
\affiliation{\USAL}
\author{Alessio \surname{Lamperti}}
\affiliation{\CNR}
\author{Rom\'eo \surname{Juge}}
\affiliation{\spinion}
\author{Dafin\'e \surname{Ravelosona}}
\email{dafine.ravelosona@c2n.upsaclay.fr}
\affiliation{\spinion}
\affiliation{\nanosciences}

\date{\today}

\begin{abstract}
Disorder in ultra-thin magnetic films can significantly hinder domain wall motion. One of the main issues on the path towards efficient domain wall based devices remains the characterization of the pinning landscape at the nanoscale. In this paper, we study domain wall motion in W/CoFeB/MgO thin films with perpendicular magnetic anisotropy crystallized by annealing at 400$^{\circ}$C and a process based on He$^{+}$ irradiation combined with elevated temperatures. Magnetic properties are similar for the whole series of samples, while the magnetic domain wall mobility is critically improved in the irradiated samples. By using an analytical model to extract nanoscale pinning parameters, we reveal important variations in the disorder of the crystallized samples. This work offers a unique opportunity to selectively analyze the effects of disorder on the domain wall dynamics, without the contribution of changes in the magnetic properties. Our results highlight the importance of evaluating the nanoscale pinning parameters of the material when designing devices based on domain wall motion, which in return can be a powerful tool to probe the disorder in ultra-thin magnetic films. 
\end{abstract}

\maketitle

\section{Introduction}
Spintronic devices based on magnetic domain wall (DW) motion are excellent candidates for logic \cite{luo_current-driven_2020, currivan_low_2012, currivan-incorvia_logic_2016, alamdar_domain_2021, raymenants_nanoscale_2021}, data-storage \cite{parkin_magnetic_2008, zhao_compact_2011} and neuromorphic devices \cite{lequeux_magnetic_2016, liu_domain_2021}. In these devices, the DWs can be efficiently driven by a current through spin-transfer torque \cite{brataas_current-induced_2012} or spin-orbit torque (SOT) \cite{manchon_current-induced_2019, emori_current-driven_2013} above a current density threshold. The latter is related to pinning of the DW due to disorder in the material, which limits the efficiency of DW motion based devices. This disorder usually takes the form of spatial variation of magnetic properties due to interface roughness, intermixing, crystalline texture, defects in the material, or grain boundaries. As a result, DWs are depinned only above a threshold force ($f_\mathrm{d}$), below which their motion is thermally activated. 

The behavior of DWs below and close to the depinning force is universal \cite{lemerle_domain_1998, chauve_creep_2000, jeudy_universal_2016, diaz_pardo_universal_2017}, and can be described using statistical models \cite{kolton_dynamics_2006, kolton_creep_2009, ferrero_spatiotemporal_2017, ferrero_creep_2021}. At driving forces below the threshold ($f < f_\mathrm{d}$), often called the creep regime, the thermally activated DW motion can be described by the interplay between the elastic energy of the DW and pinning interactions. The velocity of the DW follows an Arrhenius law $v \sim e^{-\Delta E/k_B T}$, where $k_B$ is the Boltzmann constant and $\Delta E$ the effective energy barrier, whose evaluation is far from trivial. The energy barrier scales with the driving force $\Delta E \sim f^{-\mu}$, where $\mu$ is the universal creep exponent, which depends only on the dimensionality of the system \cite{jeudy_universal_2016}. Around the depinning transition ($f \rightarrow f_d$), the velocity follows a power law with both the temperature and driving force via $v \sim T^\psi$ and $v \sim (f-f_d)^\beta$ respectively. $\beta$ and $\psi$ are also universal critical exponents \cite{diaz_pardo_universal_2017}.

The optimization of DW motion in devices is usually based on improving the magnetic parameters such as damping, Dzyaloshinskii-Moriya interaction (DMI) or effective anisotropy, and their homogeneity which reflects the structural quality of the film. Another indication of efficient DW motion is the value of the so-called depinning field $H_\mathrm{dep}$ that can be extracted from the depinning transition described above. However, these different magnetic parameters and their contributions to the underlying physics are not well understood. Several techniques based on disorder modifications have already been shown to impact the DW motion, such as varying material growth conditions \cite{lavrijsen2015asymmetric}, layer thickness \cite{metaxas_creep_2007, quinteros_correlation_2018}, or interface engineering by He$^{+}$ irradiation \cite{herrera_diez_enhancement_2019, zhao_enhancing_2019}. However, these techniques modify both the magnetic parameters and the disorder landscape, making it difficult to extract the effect of the latter on DW motion. Mostly numerical investigations studying DW motion in different crystalline environments have been carried out \cite{min_effects_2010, leliaert_current-driven_2014, voto_effects_2016, caballero_magnetic_2018}, but experimental investigations without the influence of changing magnetic parameters are still lacking. 

Understanding and optimizing DW motion in devices thus requires characterizing the disorder at the nanoscale. To this end, we studied DW motion in W/CoFeB/MgO ultra-thin films. Such films are archetype materials for DW based SOT devices, combining perpendicular magnetic anisotropy (PMA), DMI and low magnetic damping \cite{liu2011ferromagnetic} with high tunnelling magnetoresistance when used in a magnetic tunnel junction \cite{ikeda2010perpendicular}. Specifically, we studied the DW dynamics measured in samples crystallized by either pure annealing at 400$^{\circ}$C or He$^{+}$ irradiation combined with elevated temperature and with varying He$^{+}$ fluence. Using a combination of Kerr microscopy, magnetic characterizations, micromagnetic simulations and analytic modelling \cite{gehanne_strength_2020}, we can extract information about the nanoscale pinning landscape, such as the height of the average pinning energy barrier, the characteristic pinning range and pinning force, as well as the mean distance between pinning sites. In particular, by taking into account the minor differences in the magnetic parameters of the different samples, we show that the discrepancy observed in the DW dynamics in all regimes can only be ascribed to variations in the nanoscale pinning landscape, information that is out of reach of macroscopic measurement techniques.

This paper is divided into two main parts. In Section \ref{section:characterization}, we discuss the magnetic and structural characterization of the W/CoFeB/MgO samples of this study. In Section \ref{section:domainwallmotion}, we study the DW motion in the creep, depinning, and flow regimes, the microscopic pinning parameters extracted from the analytical model, and the interplay between the domain wall velocity in the flow regime and intrinsic material disorder.

\section{Sample Characterization} \label{section:characterization}
\begin{figure}[t]
\centering
\includegraphics[width=.35\textwidth]{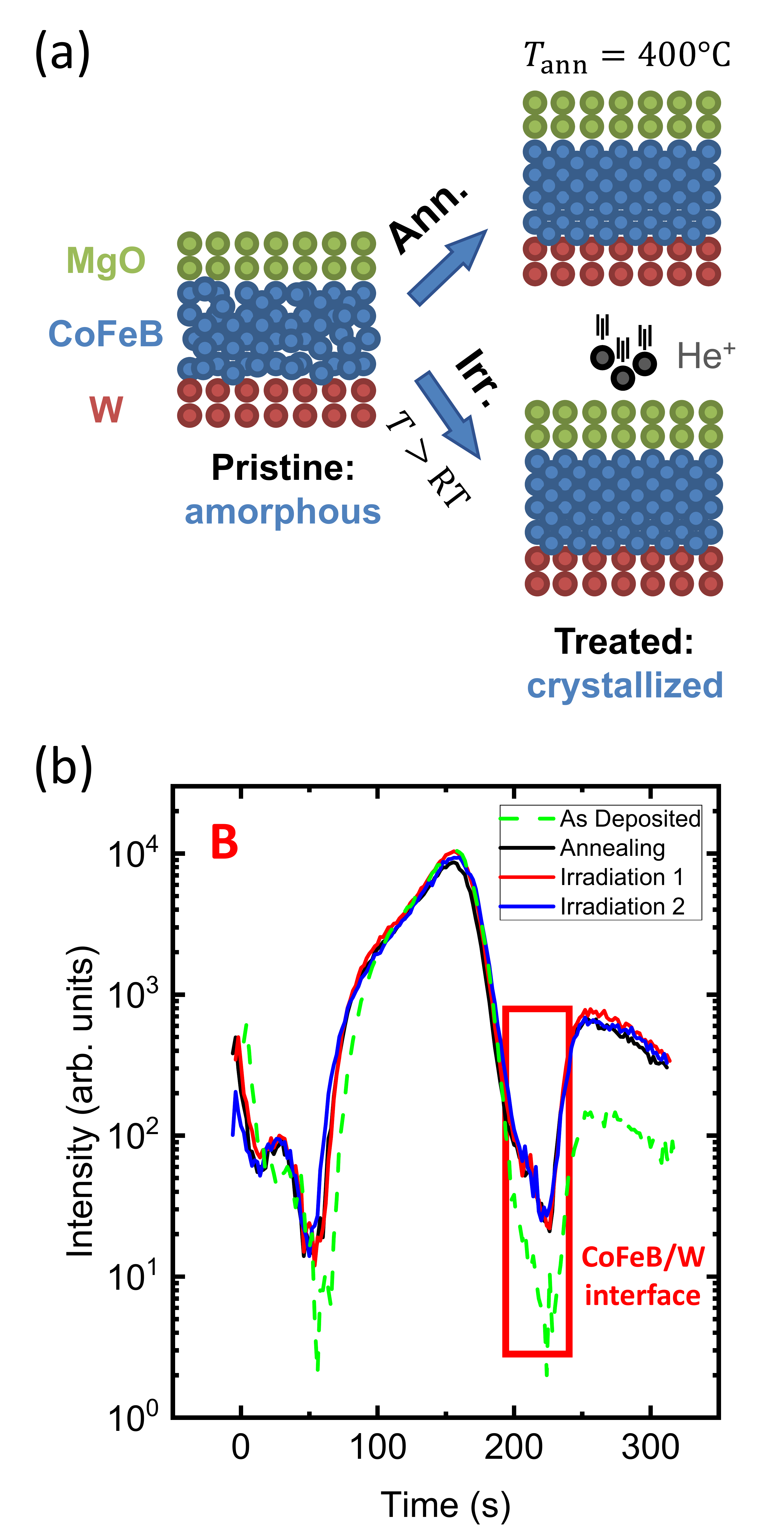}
\caption{Crystallization of magnetic layers. (a) Cartoon showing the crystallization of an amorphous CoFeB layer in a W/CoFeB/MgO stack upon He$^{+}$ irradiation at $T > \mathrm{RT}$. (b) ToF-SIMS profile of B in the amorphous sample (green) and in the three crystallized samples (black, red and blue). The red box highlights the W/CoFeB interface. Profile obtained by sputtering with O$^{2+}$ ions at 1 keV.}
\label{fig:crystallization}
\end{figure}

We investigated W(4)/Co$_\mathrm{20}$Fe$_\mathrm{60}$B$_\mathrm{20}$(1)/MgO(2)/Ta(3) (thickness in nanometers) ultra-thin films with an amorphous CoFeB layer, grown by a Singulus Rotaris sputtering tool on a 200 mm Si/SiO$_\mathrm{2}$ substrate. Three differently crystallized samples were studied, one annealed at 400$^{o}$C for 1h, and two obtained through a process based on He$^{+}$ irradiation at elevated temperatures \cite{devolder_irradiation-induced_2013}, as schematically shown in Figure \ref{fig:crystallization}a. We applied two different He$^{+}$ fluences $F_\mathrm{1} = 2 \times 10^{15}$ ions/cm$^{2}$ and $F_\mathrm{2} = 1.4 \times 10^{15}$ ions/cm$^{2}$ at 15 keV. At these irradiation parameters, no etching occurs and only short-range inter-atomic displacements are generated as a result of the irradiation. The irradiation treatments are hereafter denoted by irradiation treatment 1 and 2.

To verify the crystallization and interface ordering, we performed out-of-plane (OOP) alternating gradient magnetometry (AGM) measurements on the as-deposited and crystallized samples. The details and results of these measurements can be found in section S1 of the Supplemental Material \cite{supplemental}. The as-deposited samples show in-plane anisotropy, whereas the three treated samples all have PMA, with square hysteresis loops. Furthermore, we performed time-of-flight secondary ion mass spectroscopy (ToF-SIMS) measurements to study the evolution of the B concentration upon treatment. Figure \ref{fig:crystallization}b shows the B concentration for each sample as a function of the sputtering time, which is related to the depth. The position of the CoFeB/W interface is highlighted with a red box. All the crystallized samples show similar behavior with a significantly larger B concentration in the W compared to the as-deposited sample. The B-out diffusion from the CoFeB to the W layer and square hysteresis loops (see Section S1 in the Supplemental Material \cite{supplemental}) in the treated samples are a strong indication of CoFe crystallization \cite{greer_observation_2012, pellegren_thickness_2015}.

The magnetic properties of the crystallized samples have been investigated further through superconducting quantum interference device with vibrating sample magnetometry to measure the saturation magnetization $M_\mathrm{s}$, ferromagnetic resonance (FMR) to measure the effective anisotropy $K_\mathrm{eff}$ and the Gilbert damping $\alpha$, and Brillouin light scattering (BLS)  to measure the strength of the DMI $D$. More in-depth information on the magnetic characterization measurements is given in Section S1 of the Supplemental Material \cite{supplemental}. The results are listed in Table \ref{tab:properties}.

\begin{table}[t]
\centering
\caption{Macroscopic magnetic parameters for each sample. The table contains values for the saturation magnetization $M_\mathrm{s}$, the effective anisotropy $K_\mathrm{eff}$, the uniaxial anisotropy $K_\mathrm{u}$, the Gilbert damping $\alpha$, the DMI strength $D$ and the ratio of $D/M_\mathrm{s}$.}
\begin{tabular}{c c c c}
\hline \hline
 & Ann. & Irr. 1 & Irr. 2 \\
\hline
$M_\mathrm{s}$ (MA/m) & 1.25 $\pm$ 0.13 & 1.55 $\pm$ 0.16 & 1.60 $\pm$ 0.16 \\
$K_\mathrm{eff}$ (MJ/m$^{3}$) & 0.21 $\pm$ 0.03 & 0.20 $\pm$ 0.02 & 0.23 $\pm$ 0.02\\
$K_\mathrm{u}$ (MJ/m$^{3}$) & 1.2 $\pm$ 0.2 & 1.7 $\pm$ 0.3 & 1.8 $\pm$ 0.3\\
$\alpha$ (10$^{-3}$) & 19 $\pm$ 2 & 17 $\pm$ 1 & 18 $\pm$ 1\\
$D$ (mJ/m$^{2}$) & 0.18 $\pm$ 0.02 & 0.16 $\pm$ 0.02 & 0.25 $\pm$ 0.04\\
$D/M_\mathrm{s}$ (nJ/Am) & 0.14 $\pm$ 0.03 & 0.11 $\pm$ 0.03 & 0.16 $\pm$ 0.03\\
\hline
\end{tabular}
\label{tab:properties}
\end{table}

Overall, the three samples have very similar magnetic characteristics. Taking into account the error bars, all parameters are within 10\% of variation and similar to what has been reported for crystallized W/CoFeB/MgO thin films \cite{soucaille_probing_2016}. The annealed sample has a lower $M_\mathrm{s}$ compared to the two irradiated samples, maybe due to more intermixing at the W/CoFeB interface, since W/Fe and W/Co alloys are mostly paramagnetic \cite{nicolenco_mapping_2018}. Furthermore, the sample with irradiation treatment 2 has larger DMI compared to the other two samples. Since the DMI in W/CoFeB/MgO originates mainly from the W/CoFeB interface \cite{kuepferling_measuring_2020}, a larger DMI could suggest smoother interfaces \cite{zimmermann_dzyaloshinskii-moriya_2018}. The irradiated samples also have a larger $K_\mathrm{u}$, which could suggest that they have a smoother CoFeB/MgO interface \cite{dieny_perpendicular_2017}. These minor differences in magnetic parameters will be taken into account for the remainder of this paper. In particular, we will show that these variations in magnetic parameters are not sufficient to explain the different DW dynamics observed in the three samples, and that an investigation of the nanoscale pinning parameters is necessary to explain the DW dynamics. For a more in-depth discussion on the effect of $M_\mathrm{s}$ on the results in this work, we refer to section S2 of the Supplemental Material \cite{supplemental}.     

\section{Domain Wall Motion}\label{section:domainwallmotion}
We now study the DW motion in order to get insight into the microscopic structure of the samples. To do so, we investigate both the expansion profile and the velocity of the DW as a function of the OOP applied magnetic field using polar Kerr microscopy. By using an analytical model from G\'ehanne \textit{et al.} \cite{gehanne_strength_2020} for the DW motion in the creep regime and micromagnetic simulations in the flow regime, we can extract more detailed information about the microscopic structure and pinning properties of the samples. We refer to section S3 in the Supplemental Material for a more in-depth description of the DW motion measurement techniques used for the experiments in this section \cite{supplemental}.

\subsection{Creep and depinning}
\begin{figure*}[t]
\centering
\includegraphics[width=.8\textwidth]{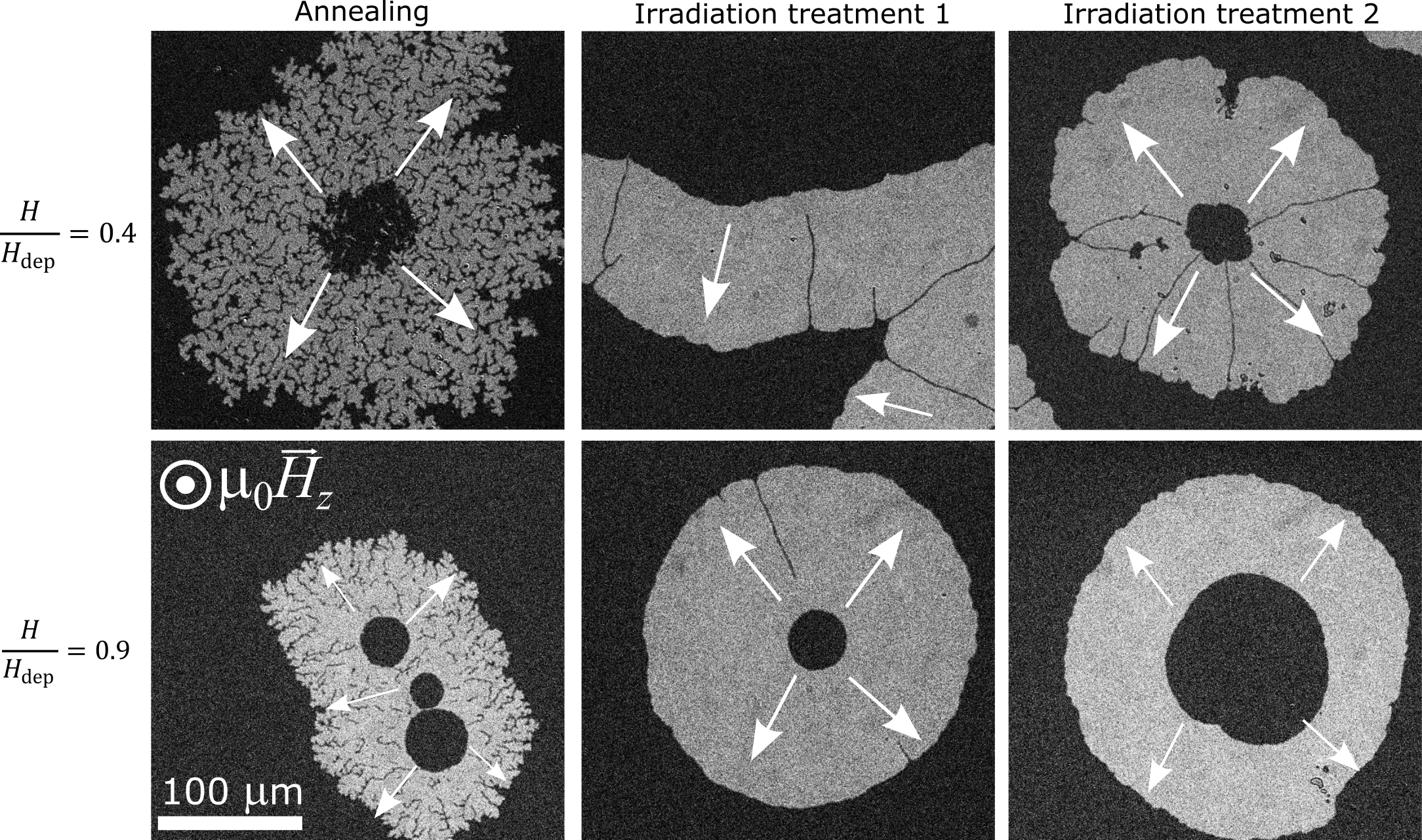}
\caption{DW expansion images for the three samples investigated in the creep regime at $H/H_\mathrm{dep} = 0.4$ and close to the depinning transition at $H/H_\mathrm{dep} = 0.9$. Images taken with (1 pulse of 4 seconds at 1.83 mT, 4 pulses of 10 $\mu$s at 4.45 mT), (2 pulses of 10 ms at 1.91 mT, 6 pulses of 10 $\mu$s at 4.22 mT), and (4 pulses of 10 ms at 2.20 mT, 2 pulses of 10 $\mu$s at 5.40 mT) for the three samples at ($H/H_\mathrm{dep} = 0.4$, $H/H_\mathrm{dep} = 0.9$) respectively. All images are differential images, where the light-grey areas indicate the area over which the DW has moved during the OOP magnetic field pulse. The white arrows denote the direction of movement of the DW.}
\label{fig:images}
\end{figure*}

Figure \ref{fig:images} shows differential Kerr microscopy images for each sample at two different values of the reduced magnetic field; in the creep regime at $H/H_\mathrm{dep} = 0.4$ and  closer to the depinning transition at $H/H_\mathrm{dep} = 0.9$. A striking result is that in the annealed sample the DW expansion exhibits fractal-like features, whereas the irradiated samples show bubble-like expansion with overall lower jaggedness. This difference in domain structure is an indication of stronger pinning effects in the annealed sample. Furthermore, smoother domain walls have been linked to smoother interfaces \cite{magni_key_2022}. We can also observe a smoothing of the DW profile when increasing the field towards the depinning transition, which is consistent with current models highlighting a reduction of the roughness amplitude at higher driving forces \cite{cortes_burgos_field-dependent_2021}.  

\begin{figure}[t]
\centering
\includegraphics[width=.5\textwidth]{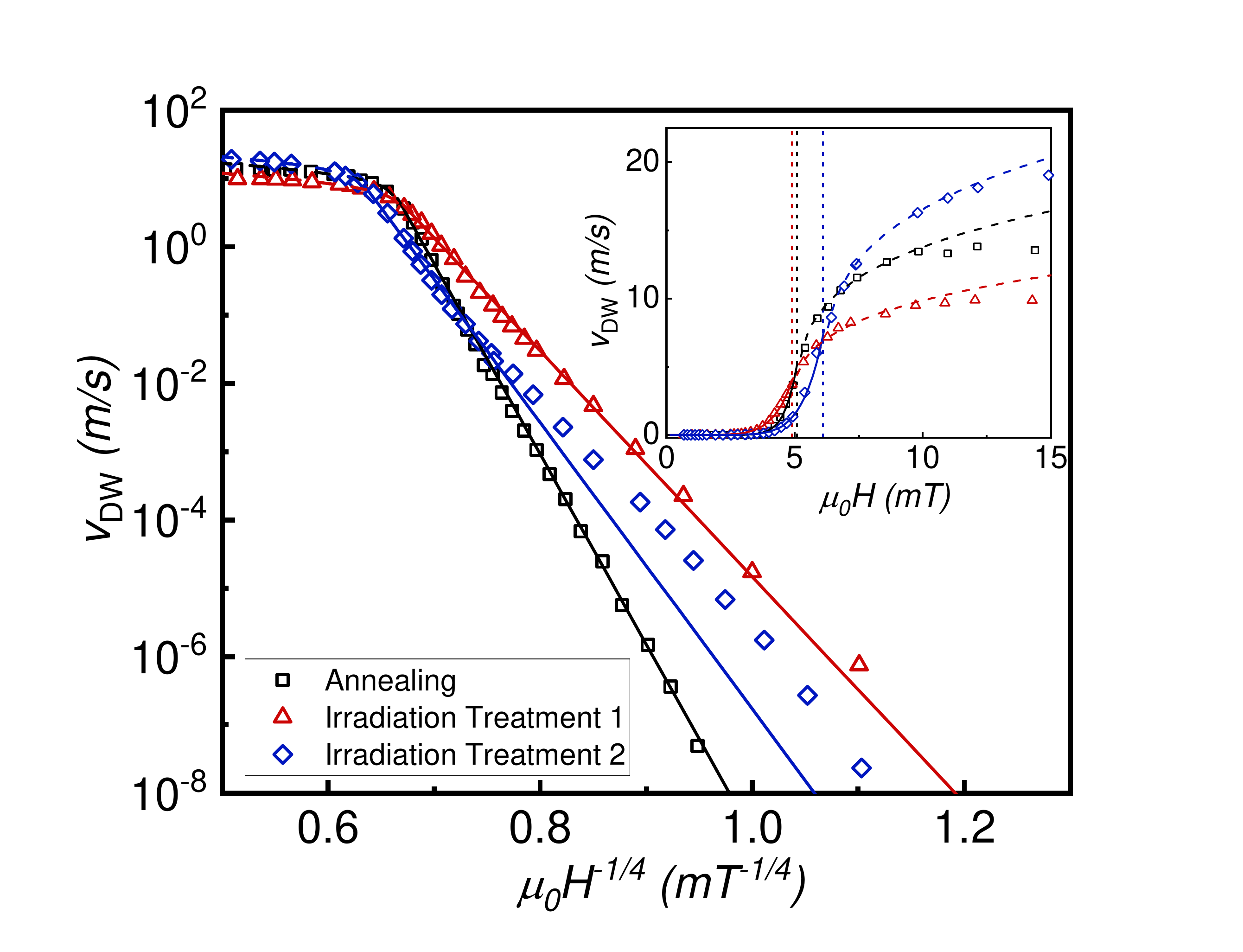}
\caption{DW velocity as a function of the OOP applied magnetic field strength, rescaled to show creep law behavior. The solid lines are fits in the creep regime, and the dashed lines are fits in the depinning regime. The inset is a representation with linear axes around the depinning transition. The dotted vertical lines in the inset denote the depinning fields $H_\mathrm{dep}$.}
\label{fig:velocitycreep}
\end{figure}

Figure \ref{fig:velocitycreep} shows the DW velocity as a function of the applied magnetic field in the creep regime. We observe a strong variation of the DW velocity of several orders of magnitude, with the irradiated samples having a larger velocity than the annealed sample. The velocity difference is as high as 4 orders of magnitude at 0.94 (mT)$^{-1/4}$ (1.28 mT). Such higher DW velocity is consistent with easier DW motion as revealed by the bubble like expansion.

To go further, we now use domain wall velocity as a function of an OOP applied magnetic field to extract more information about the domain wall motion mechanism in our samples. In the creep and depinning regimes, the velocity is given by \cite{jeudy_universal_2016, diaz_pardo_universal_2017}:

\begin{align}
v(H) = \begin{cases} v\left(H_\mathrm{dep}\right)\exp\left(-\frac{\Delta E}{k_\mathrm{B} T}\right)\mbox{,} & \mbox{c.: } H < H_\mathrm{dep} \\ \frac{v\left(H_\mathrm{dep}\right)}{x_\mathrm{0}}\left(\frac{T_\mathrm{d}}{T}\right)^{\psi}\left(\frac{H-H_\mathrm{dep}}{H_\mathrm{dep}}\right)^{\beta}\mbox{,} & \mbox{d.: } H \gtrsim H_\mathrm{dep}\mbox{,} \end{cases}
\label{eq:creepanddepinning}
\end{align}

where $\Delta E = k_\mathrm{B}T_\mathrm{d}\left(\left(H/H_\mathrm{dep}\right)^{-\mu}-1\right)$ is the creep energy barrier. Here we assume the quenched Edwards-Wilkinson universality class for creep and depinning \cite{edwards_surface_1982}, with the critical exponents fixed at $\mu = 1/4$, $\beta=0.25$ and $\psi = 0.15$, and the universal constant taken at $x_\mathrm{0} = 0.65$. The non-universal (i.e. material dependent) parameters are the depinning field $H_\mathrm{dep}$ and its corresponding velocity $v\left(H_\mathrm{dep}\right)$, as well as the effective pinning barrier height $k_\mathrm{B}T_\mathrm{d}$, with $T_\mathrm{d}$ the depinning temperature.

We used the expressions in Eq. \ref{eq:creepanddepinning} to fit the domain wall velocity reported in Figure \ref{fig:velocitycreep}. These fits are represented by solid and dashed lines for the creep and depinning regimes, respectively. The vertical dotted line in the inset denotes the depinning field $H_\mathrm{dep}$. Note that we performed the fit close to the depinning transition. Due to the change in slope for the sample with irradiation treatment 2, the fit does not work deeper in the creep regime, and the following analysis is thus only valid close to the depinning regime. The other samples do not have this problem. The change in slope has been seen before, but its origin is still unknown \cite{caballero_excess_2017}. The resulting values for the non-universal parameters are reported in Table \ref{tab:pinning}.

\begin{table}[t]
\centering
\caption{Creep and depinning parameters extracted from the DW velocity curves using Eq. \ref{eq:creepanddepinning}, domain wall parameters, microscopic pinning parameters extracted with Eqs. \ref{eq:xi} and \ref{eq:fpin}, and other disorder parameters extracted from FMR and BLS linewidth measurements. The table reports values for the depinning field $H_\mathrm{dep}$, the pinning barrier height $k_\mathrm{B}T_\mathrm{d}$, the domain wall width $\Delta$, domain wall energy density times the thickness $\sigma \times t$, the characteristic pinning range $\xi$, the depinning force $f_\mathrm{pin}$, the average distance between pinning sites $\frac{1}{\sqrt{n}}$ based on the scaling proposed in ref \cite{gehanne_strength_2020}, the inhomogeneous broadening $\mu_\mathrm{0}\Delta H_\mathrm{0}$ as extracted by FMR, and both the raw effective anisotropy field variation $\Delta H_{K_\mathrm{eff}}$ and reduced $\Delta H_{K_\mathrm{eff}}/H_{K_\mathrm{eff}}$ as extracted by BLS. Note that the reduced effective anisotropy field value has been calculated with the anisotropy fields as extracted from BLS measurements.}
\begin{tabular}{c c c c}
\hline \hline
& Ann. & Irr. 1 & Irr. 2\\ \hline
\multicolumn{4}{c}{Creep and depinning parameters}\\
$\mu_\mathrm{0} H_\mathrm{dep}$ (mT) & 5.1 $\pm$ 0.1 & 4.9 $\pm$ 0.1 & 6.1 $\pm$ 0.1\\
$k_\mathrm{B}T_\mathrm{d}$ (eV) & 1.12 $\pm$ 0.09 & 0.69 $\pm$ 0.09 & 0.78 $\pm$ 0.09\\
\multicolumn{4}{c}{Domain wall parameters}\\
$\Delta$ (nm) & 8.4 $\pm$ 0.4 & 8.6 $\pm$ 0.4 & 8.1 $\pm$ 0.4\\
$\sigma \times t$ (pJ/m) & 7.2 $\pm$ 0.4 & 6.9 $\pm$ 0.3 & 7.4 $\pm$ 0.4\\
\multicolumn{4}{c}{Microscopic pinning parameters}\\
$\xi$ (nm) & 69 $\pm$ 5 & 47 $\pm$ 5 & 47 $\pm$ 4\\
$20 \times f_\mathrm{pin}$ (pJ/m) & 5.2 $\pm$ 0.3 & 6.4 $\pm$ 0.5 & 7.9 $\pm$ 0.6\\
$\frac{1}{\sqrt{n}}$ (nm) & 17 $\pm$ 2 & 13 $\pm$ 2 & 12 $\pm$ 2\\
\multicolumn{4}{c}{Other disorder parameters}\\
$\mu_\mathrm{0}\Delta H_\mathrm{0}$ (mT) & 36 $\pm$ 3 & 46 $\pm$ 1 & 30 $\pm$ 1 \\
$\mu_\mathrm{0}\Delta H_{K\mathrm{eff}}$ (mT) & 24 $\pm$ 1 & 16 $\pm$ 3 & 20 $\pm$ 1\\
$\frac{\Delta H_{K_\mathrm{eff}}}{H_{K_\mathrm{eff}}}$ (\%) & 5.6 $\pm$ 0.4 & 5.3 $\pm$ 1.2 & 5.3 $\pm$ 0.5\\
\hline
\end{tabular}
\label{tab:pinning}
\end{table}

We find that the irradiated samples have a significantly lower $k_\mathrm{B}T_\mathrm{d}$ than the annealed sample. In particular, $k_\mathrm{B}T_\mathrm{d}$ is 38\% lower for the sample with irradiation treatment 1, and 31\% lower for the sample with irradiation treatment 2 compared to the annealed sample. A larger pinning barrier for the annealed sample is consistent with stronger pinning sites, as seen in Figure \ref{fig:images}. The depinning field remains fairly constant, where the sample with irradiation treatment 2 has a slightly larger $\mu_\mathrm{0}H_\mathrm{dep}$ than the others. While typically a larger depinning field is a signature of stronger pinning, our results indicate that it is not the sole contributor to the pinning strength, as can clearly be seen in Figure \ref{fig:images} for the annealed sample.

\subsection{The role of microscopic pinning parameters}
\begin{figure}[t]
\centering
\includegraphics[width=.25\textwidth]{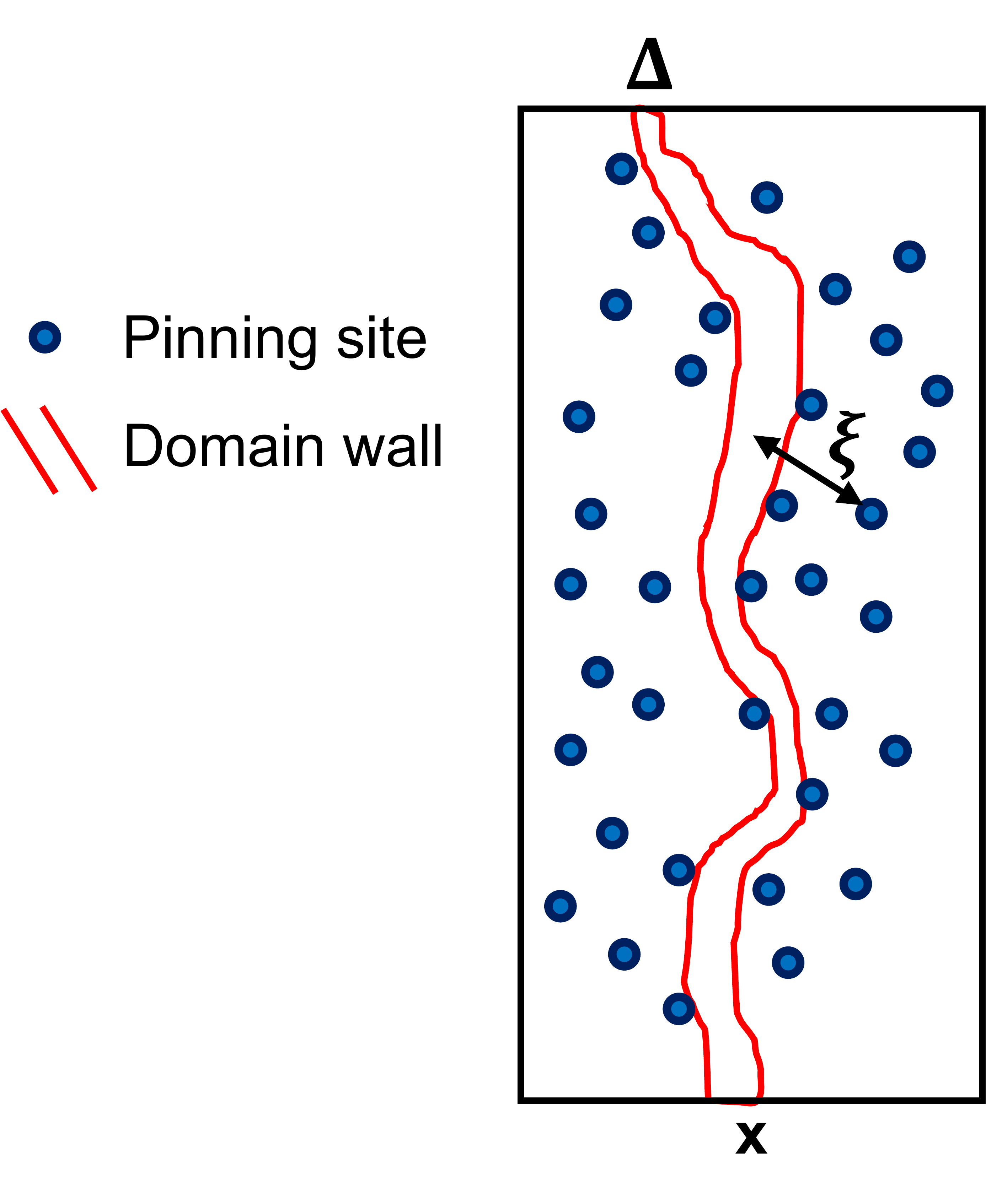}
\caption{Schematic representation of a DW with width $\Delta$ surrounded by pinning sites acting on it. $\xi$ denotes the range of interaction between the pinning sites and the DW.}
\label{fig:pinningcartoon}
\end{figure}

With the magnetic characterization and non-universal DW motion parameters extracted from Eq. \ref{eq:creepanddepinning}, we can now use the analytical model to extract the microscopic pinning parameters \cite{gehanne_strength_2020}. The model considers a DW with width $\Delta = \sqrt{A/K_\mathrm{eff}}$ (with $A$ the exchange stiffness) and surface energy density $\sigma$ (in J/m$^{2}$) interacting with the surrounding pinning sites, as schematically shown in Figure \ref{fig:pinningcartoon}. In order to calculate $\sigma$ and $\Delta$, we assume $A =$ 15 pJ/m \cite{zhao_enhancing_2019}.

We can then obtain the characteristic range $\xi$ (nm) and strength $f_\mathrm{pin}$ (pJ/m) of the DW-defect interaction by using the following expressions:

\begin{align}
\xi &\sim \left[\left(k_\mathrm{B}T_\mathrm{d}\right)^{2}/\left(2\mu_\mathrm{0}H_\mathrm{dep}M_\mathrm{s}\sigma t^{2}\right)\right]^{1/3}, \label{eq:xi}\\
f_\mathrm{pin} &\sim \frac{1}{\sqrt{n}\xi}\sqrt{2\mu_\mathrm{0}H_\mathrm{dep}M_\mathrm{s}tk_\mathrm{B}T_\mathrm{d}}.
\label{eq:fpin}
\end{align}

Here $t$ is the thickness of the magnetic film, and $\frac{1}{\sqrt{n}}$ the average distance between pinning sites, with $n$ the pinning site density per unit area. We assume that the nominal thickness $t =$ 1 nm of the magnetic layer does not change, since slight variations in dead layers are already taken into account in the $M_\mathrm{s}$ value. The value of $\frac{1}{\sqrt{n}}$ is based on the scaling between $\xi$, $f_\mathrm{pin}$, $\Delta$ and $\sigma \times t$ as proposed in ref \cite{gehanne_strength_2020}, whose values are all reported in table \ref{tab:pinning}. 

As proposed in ref \cite{gehanne_strength_2020}, $f_\mathrm{pin}$ should be proportional to $\sigma \times t$. When assuming $f_\mathrm{pin} \approx \sigma t/20$ and that the three samples have the same average distance between pinning sites, we find $\frac{1}{\sqrt{n}} = 12$ nm, which is close to the typical grain size in polycrystalline films \cite{zeissler_pinning_2017}. The agreement between 20$f_\mathrm{pin}$ and $\sigma \times t$ is good for the irradiated sample, but not for the annealed sample. For the annealed sample, one way to obtain the proportionality is to increase the value of $\frac{1}{\sqrt{n}}$, which suggests that the microscopic structure of the three samples is indeed not the same. For each sample, the exact values of $\frac{1}{\sqrt{n}}$ obtained with this method are also reported in Table \ref{tab:pinning}. Analysis of the microscopic pinning parameters thus shows that the annealed sample, while having strong pinning sites, has an overall larger mean spacing between pinning sites compared to the irradiated samples. This is further strengthened by the strongly reduced characteristic pinning range $\xi$ for the irradiated samples. The decreased spacing between pinning sites and characteristic pinning range could potentially be an effect caused by a reduction of grain size in the irradiated samples.

Table \ref{tab:pinning} also contains other parameters which are often used to quantify disorder in materials, such as the the inhomogeneous broadening $\mu_\mathrm{0}\Delta H_\mathrm{0}$ and effective anisotropy field variations $\Delta H_{K_\mathrm{eff}}/H_{K_\mathrm{eff}}$, which can be extracted from FMR and BLS linewidth experiments (see section S1 of the Supplemental Material \cite{supplemental}). The inhomogeneous broadening from FMR measurements shows significant variation between the three samples, with the sample with irradiation treatment 1 having the largest broadening. While this is an indication that the disorder is different between the three samples, no obvious trend between the broadening and domain wall motion is observed. Scale can be a potential reason for the discrepancy between the linewidth measurements and domain wall motion, since domain wall motion is typically measured over a much shorter length scale compared to FMR.

Contrary to the broadening measured in FMR, the raw effective anisotropy field variations measured with BLS show small changes that are consistent with the reduction in $k_\mathrm{B}T_\mathrm{d}$ and the increase in creep velocity showed in Figure \ref{fig:velocitycreep}, where the samples with lower $\mu_\mathrm{0}\Delta H_{K\mathrm{eff}}$ also have faster creep motion and a lower $k_\mathrm{B}T_\mathrm{d}$. However, when looking at the reduced effective anisotropy field variation $\Delta H_{K_\mathrm{eff}}/H_{K_\mathrm{eff}}$ the effect disappears. Furthermore, the values obtained from FMR and BLS do not agree, which we discuss in more detail in section S4 of the Supplemental Material \cite{supplemental}.

\subsection{Disorder and the flow regime}
\begin{figure}[t]
\centering
\includegraphics[width=.46\textwidth]{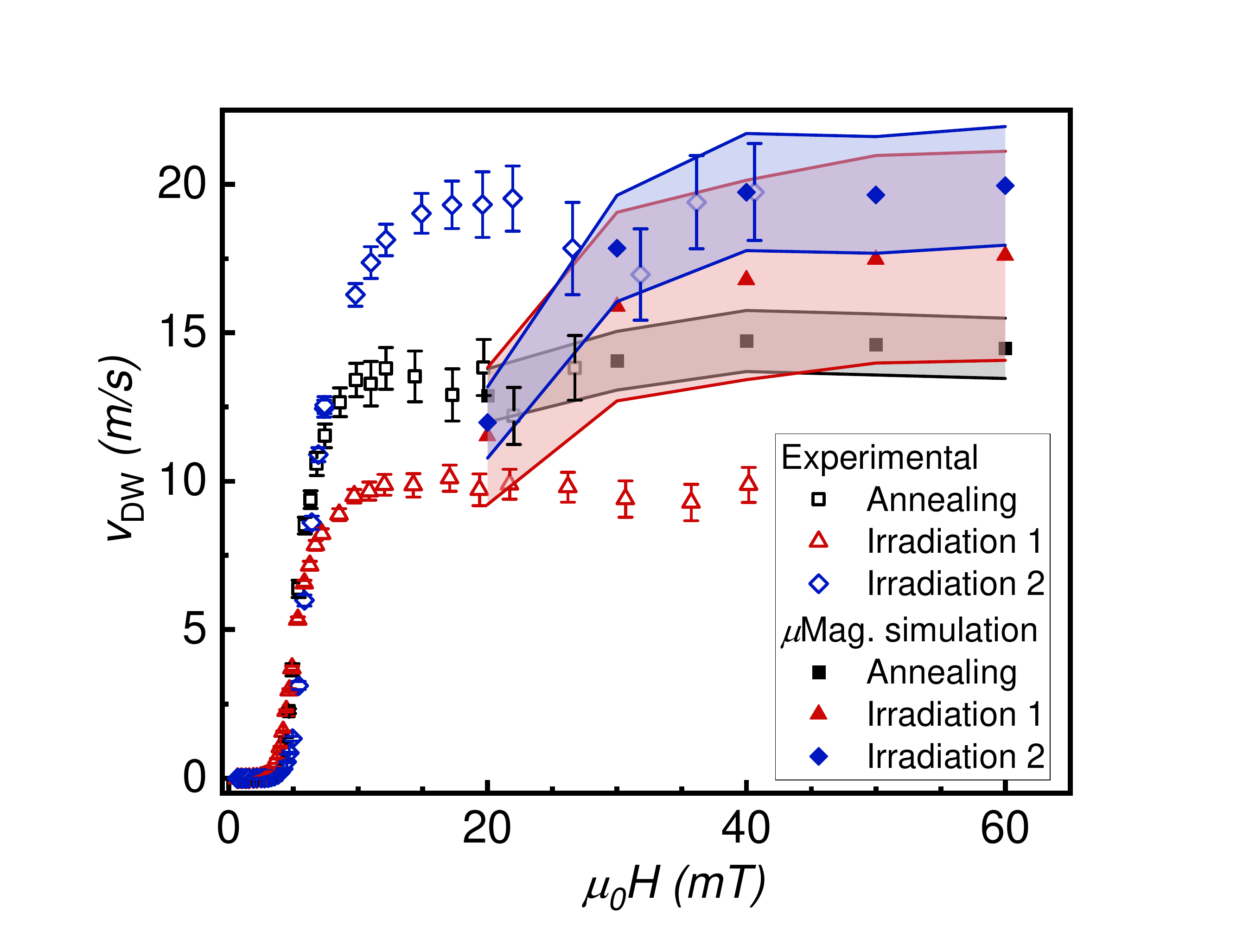}
\caption{Domain wall velocity as a function of the OOP applied magnetic field for the three samples. The figure contains experimental results (open symbols) and results from micromagnetic simulations with disorder (closed symbols). The shaded area for the micromagnetic simulation denotes variation in domain wall velocity based on the measured anisotropy variation of BLS and the characterization results of the micromagnetic simulations (see sections S1 and S5 of the Supplemental Material \cite{supplemental}).}
\label{fig:mumaxflow}
\end{figure}

We also studied the domain wall velocity in the flow regime, which is shown in Figure \ref{fig:mumaxflow} with open symbols. We see that the domain wall velocity reaches a plateau above the depinning transition. This has been attributed to precessional losses inside the DW at fields larger than the Walker breakdown \cite{garcia_magnetic_2021}, the onset of the magnetization precession \cite{coey_magnetism_2010}. We observe that the plateau lies at different velocity values about 13 m/s, 10 m/s and 19 m/s for the annealed, irradiation 1 and irradiation 2 samples, respectively. 

Recent work has shown that in disordered samples with weak DMI, the height of the velocity plateau depends on $D/M_\mathrm{s}$ \cite{garcia_magnetic_2021}. While qualitatively the height of the velocity plateau scales with $D/M_\mathrm{s}$ in our samples (see Table \ref{tab:properties}), recent simulation work in polycrystalline films has shown that the disorder can strongly influence the velocity in the precessional regime, mainly through the grain size and anisotropy variations \cite{voto_effects_2016}. 

To shed more light on the effect of disorder on the domain wall velocity in the flow regime in our samples, we performed an extensive set of micromagnetic simulations in a similar vein to ref \cite{voto_effects_2016}. The details of the micromagnetic simulations can be found in section S5 of the Supplemental Material \cite{supplemental}. We simulated domain wall velocity at different magnetic fields in the precessional regime, taking into account the measured micromagnetic properties and anisotropy variation of the samples. We set the grain size constant at 10 nm - a typical value in polycrystalline films \cite{zeissler_pinning_2017}. The results of these simulations are shown with closed symbols in Figure \ref{fig:mumaxflow}. The shaded area around the simulation data denotes the change in domain wall velocity based on the error in the variation of anisotropy field that we measured with BLS (see Table \ref{tab:pinning}). For more details, we refer to sections S1, S4 and S5 of the Supplemental Material \cite{supplemental}.

Figure \ref{fig:mumaxflow} shows decent agreement between simulation and experiment for the annealed sample and the sample treated with irradiation treatment 2. However, the velocity for the sample with irradiation treatment 1 is about a factor 2 larger in the simulation compared to the experiment. Even when taking into account the error in the measured anisotropy field variation, the velocity is still too large. A possible explanation could be that the grain size of the sample treated with irradiation treatment 1 is lower than 10 nm, as even a small variation of a couple nanometers in average grain size can result in a large drop of plateau velocity \cite{voto_effects_2016} (see also section S5 of the Supplemental Material \cite{supplemental}).

\section{Discussion}\label{section:discussion}
The effect of the changes in the microscopic pinning parameters reported in Table \ref{tab:pinning} can be seen when considering the pinning free energy $\delta F_\mathrm{pin} = f_\mathrm{pin}\sqrt{nL\xi}\xi$ of the collective pinning length of the DW $L_\mathrm{c} \sim (\xi/n)^{1/3} (\sigma t/f_\mathrm{pin})^{2/3} \approx 160$ nm \cite{gehanne_strength_2020}, also called the Larkin length. We see a similar $\sim$ 30\% drop in pinning free energy due to the proportionality imposed with $k_\mathrm{B}T_\mathrm{d}$. Despite the decrease in the mean spacing of pinning sites (i.e. a higher defect density), a reduced energy barrier makes it easier for thermal excitations to induce DW motion, since the overall displacement of the DW in the creep regime is driven by the largest energy barriers \cite{kolton_creep_2009}. 

\begin{figure}[b]
\centering
\includegraphics[width=.46\textwidth]{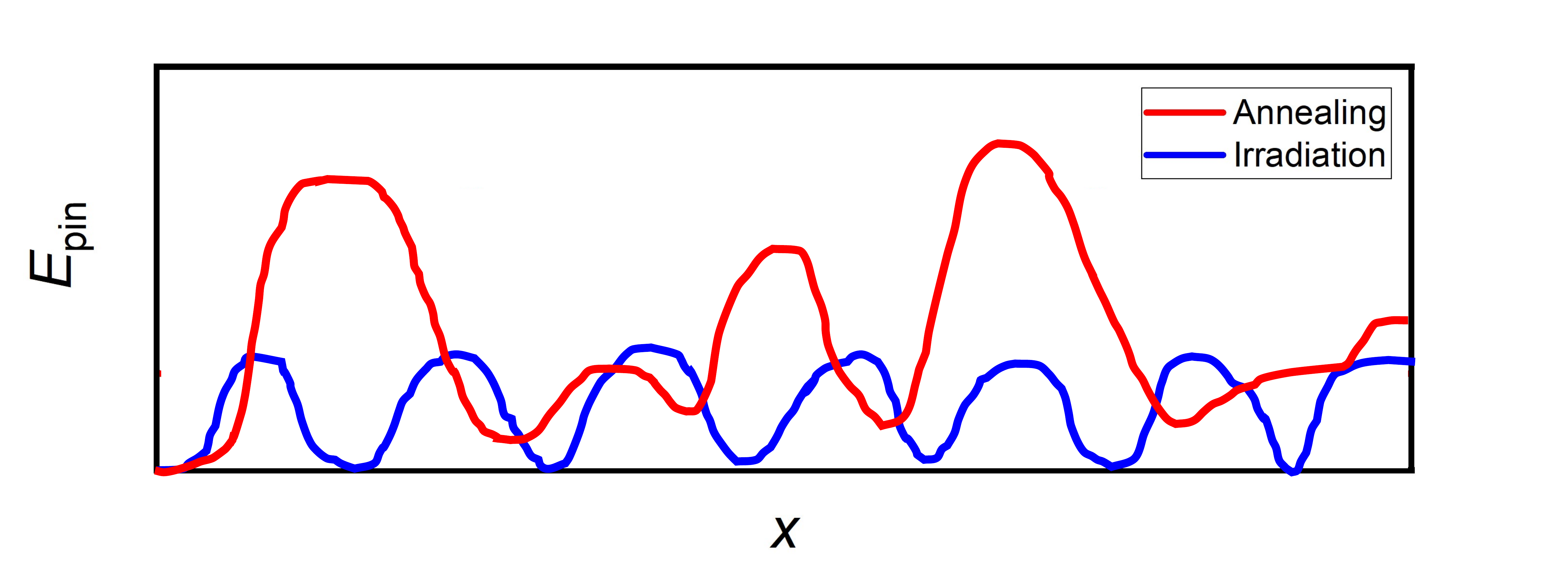}
\caption{Proposed 1D energy landscape for the annealed sample (red) and irradiated samples (blue).}
\label{fig:energylandscape}
\end{figure}

A visual description of such an pinning energy landscape is shown in Figure \ref{fig:energylandscape}. Here, two pinning landscapes are shown, in red for the annealed sample, and in blue for the irradiated samples. The lower maximum pinning energy of the irradiated samples corresponds to the reduction in pinning energy barrier, whereas the shorter spacing of the barriers represents the reduction in mean distance between pinning sites. The overall smoother landscape is a representation of the observations of smoother domain wall propagation in Figure \ref{fig:images}.

Looking beyond the depinning transition, little is still known about the interplay between intrinsic material defects and the DW velocity in the velocity plateau. While disorder does impact the domain wall velocity in the precessional regime, only the sample with irradiation treatment 1 showed indications of significantly affected disorder, which is not consistent with our results from the creep regime. Moreover, the depinning fields, which also depend on the disorder, are significantly larger in the simulation compared to the experiment. One of the main causes is the absence of temperature in the simulations, but it shows that the grain size and anisotropy fluctuations have to be tuned further in order to reproduce the experimental values of the depinning field. Furthermore, the inclusion of DMI and thickness variations \cite{gross_skyrmion_2018} can further help in reproducing both the depinning fields and the velocity in the precessional regime.

Nonetheless, we can use the results from all DW motion regimes to gain insight into the disorder. A reduction in grain size upon irradiation would explain both the reduced average distance between pinning sites, as well as some of the discrepancies in the flow regime. Using high resolution transmission electron microscopy (HR-TEM) measurements it was not possible to see any crystal structure difference due to the nanometer scale origin of disorder (interface roughness, interface intermixing, etc.).

Overall, we can identify the grain size as a critical parameter for the optimization of DW motion in polycrystalline films. The pinning energy barrier also plays a significant role in the creep regime next to the depinning field, while in the flow regime the anisotropy variation can also have a significant impact \cite{min_effects_2010}. However, more extensive simulations would be necessary to fully characterize the velocity in the flow regime and to make a link between the creep and depinning parameters and the flow regime.

Despite this fact, our results show that DW motion is an excellent tool to characterize disorder and notably provides information on the length scales involved in DW pinning, which is absent in a traditional magnetic study. Other techniques to independently access the same information are therefore required.

\section{Conclusions}
To summarize, we used DW motion to experimentally study the magnetic disorder of three differently crystallized W/CoFeB/MgO ultra-thin films. The DW roughness and velocity in the samples were observed to be significantly different in the creep and flow regimes. These differences in both regimes cannot be explained via pure magnetic characterizations, and a deeper dive into the microscopic pinning parameters is required to fully understand the domain wall motion. We used an analytical model to link changes in the creep velocity to changes in the microscopic pinning parameters through the modulated pinning energy barrier. Furthermore, we used micromagnetic simulations and the domain wall velocity in the precessional regime to obtain information about structural differences between the samples. From the extensive measurement of all DW motion regimes we find that the pinning energy barrier $k_\mathrm{B}T_\mathrm{d}$, the scaled average distance between pinning sites $\frac{1}{\sqrt{n}}$, and anisotropy variations are, on top of the depinning field $H_\mathrm{dep}$, among the critical parameters in optimizing DW motion. Furthermore, we identified the grain size as an important physical parameter, although more research is required to confirm its relation to the microscopic parameters and precessional DW velocity. Our results show that DW motion can be an efficient tool to characterize the microscopic pinning environment of samples with different crystalline structures, and that the microscopic pinning environment is a prerequisite when optimizing materials for devices based on DW motion. Not only are our results relevant for DW motion, but they could be applied to skyrmion motion as well, since it's also heavily impacted by disorder \cite{kim_current-driven_2017, legrand_room-temperature_2017, juge_current-driven_2019}.

\begin{acknowledgments}
We would like to thank Randy Dumas from Quantum Design and Fredrik Magnusson from NanOsc for performing the FMR measurements, and Thomas Hauet for performing the SQUID-VSM measurements. J.W.v.d.J., M.S., L.H.D., M.F., L.L.D, R.J. and D.R. acknowledge funding from the European Union's Framework Programme for Research and Innovation Horizon 2020 (2014-2020) under the Marie Skłodowska-Curie Grant Agreement No. 860060 (MagnEFi). L.L.D. further acknowledges Ministerio de Ciencia e Innovacion under project PID2020-117024GD-C41 and Consejeria de Educacion of Castilla y Leon under project SA114P20.
\end{acknowledgments}

\end{document}


\renewcommand{\theequation}{S\arabic{equation}}
\renewcommand{\thefigure}{S\arabic{figure}}
\renewcommand{\thetable}{S\arabic{table}}  
\renewcommand{\bibnumfmt}[1]{[S#1]}
\renewcommand{\citenumfont}[1]{S#1}
\renewcommand{\thesection}{S\arabic{section}}

\title{Revealing nanoscale disorder in W/CoFeB/MgO ultra-thin films using domain wall motion - Supplemental Material}

\author{Johannes W. \surname{van der Jagt}}
\email{gyan.vdjagt@spin-ion.com}
\affiliation{\spinion}
\affiliation{\parissaclay}
\author{Vincent \surname{Jeudy}}
\author{Andr\'e \surname{Thiaville}}
\affiliation{\LPS}
\author{Mamour \surname{Sall}}
\affiliation{\spinion}
\author{Nicolas \surname{Vernier}}
\author{Liza \surname{Herrera Diez}}
\affiliation{\nanosciences}
\author{Mohamed \surname{Belmeguenai}}
\author{Yves \surname{Roussign\'e}}
\author{Salim M. \surname{Cherif}}
\affiliation{\univparis}
\author{Mouad \surname{Fattouhi}}
\author{Luis \surname{Lopez-Diaz}}
\affiliation{\USAL}
\author{Alessio \surname{Lamperti}}
\affiliation{\CNR}
\author{Rom\'eo \surname{Juge}}
\affiliation{\spinion}
\author{Dafin\'e \surname{Ravelosona}}
\email{dafine.ravelosona@c2n.upsaclay.fr}
\affiliation{\spinion}
\affiliation{\nanosciences}

\date{\today}

\maketitle

\section{Magnetic characterization}
In this section, we will go into more detail about the magnetic characterization measurements of the three samples described in the main paper, as well as some additional measurements to complete those presented in the main paper. Specifically, we will discuss the hysteresis loops taken by MOKE, the ferromagnetic resonance (FMR), and the Brillouin light scattering (BLS) measurements.

\subsection{MOKE hysteresis loops}
\begin{figure}[hb]
\centering
\includegraphics[width=.47\textwidth]{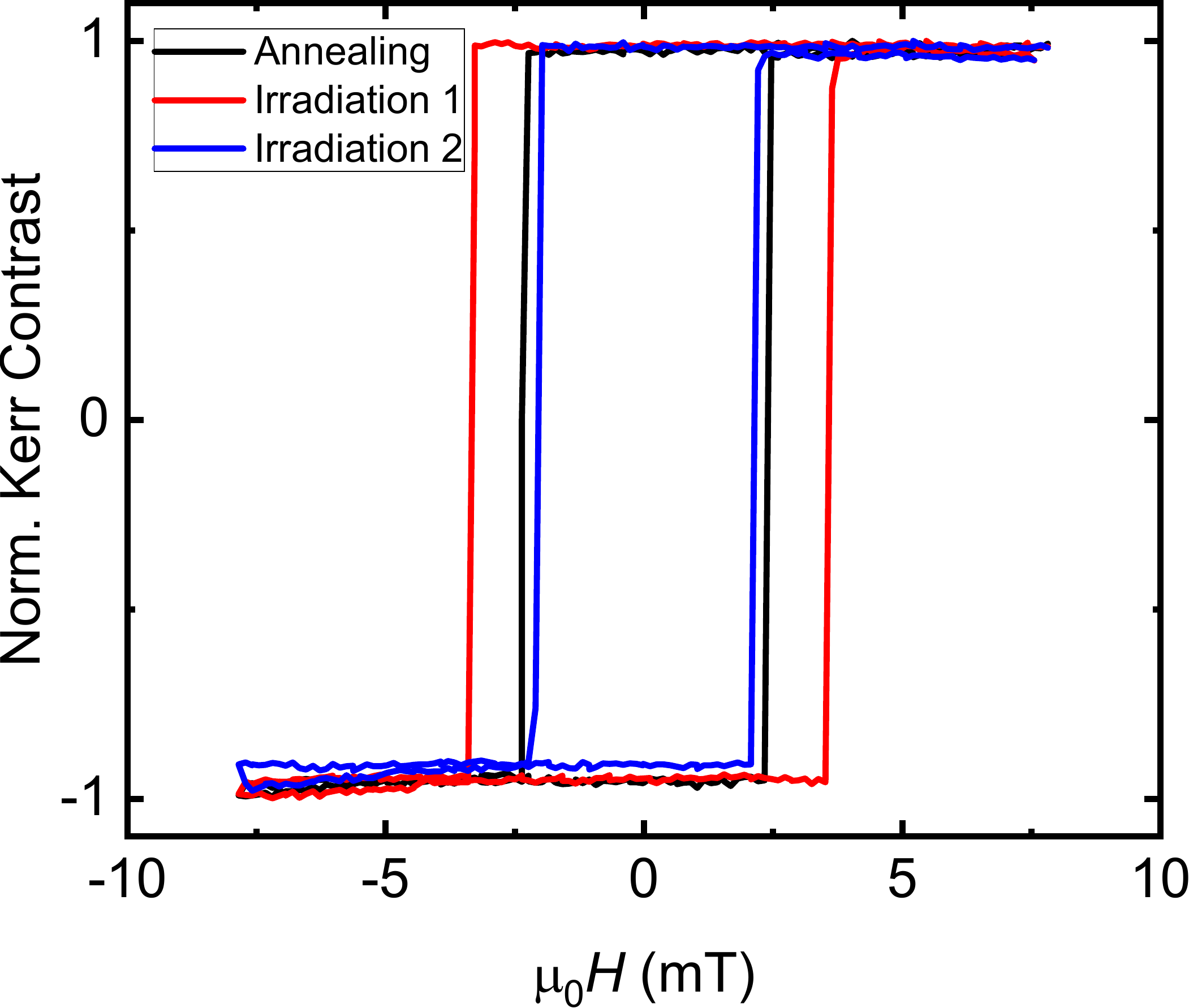}
\caption{MOKE hysteresis loops of the three samples from the main paper taken in an OOP field. Data has been treated to correct for linear offsets and slopes in the tails.}
\label{fig:hysteresisloops}
\end{figure}

We used a Zeiss Axioscope Kerr microscope to obtain the OOP hysteresis loops of the treated samples, which are shown in Figure \ref{fig:hysteresisloops}. A video is taken during a sweep of an OOP field. At each field step, the average contrast of the image is used as a measure for the magnetization. 

Figure \ref{fig:hysteresisloops} clearly shows that all three treated samples have a square hysteresis loop with similar characteristics. The sample with irradiation treatment 1 has a larger coercive field compared to the other treated samples. This could point to an increased nucleation field, or the absence of strong nucleation sites.

\subsection{Ferromagnetic resonance (FMR) measurements}
The FMR measurements in the main paper have been performed with the applied dc field in the OOP direction and with the rf field in the IP direction. Furthermore, the measurements have all been performed in field sweep mode.

The resonant frequency at different applied fields $H$ is given by the Kittel formula for spontaneously out-of-plane magnetized samples and perpendicular applied fields \cite{coey_magnetism_2010}:

\begin{align}
f = \frac{\gamma_0}{2\pi} \left(H+H_{K_\mathrm{eff}}\right),
\label{eq:KittelOOP}
\end{align}

where $\gamma_0 = \gamma \mu_0$ is the gyromagnetic ratio, and $H_{K_\mathrm{eff}} = H_{K} - M_\mathrm{s}$ is the effective anisotropy field for the perpendicular orientation.

The full width at half maximum (FWHM) linewidth of the absorption spectra can give further information about the damping effects in our samples. It is given by \cite{noel_dynamical_2019}:

\begin{align}
\Delta H = \frac{4\pi\alpha}{\gamma_0}f + \Delta H_\mathrm{0},
\label{eq:FMRlinewidth}
\end{align}

where $\alpha$ is the Gilbert damping constant, and $\Delta H_\mathrm{0}$ is the inhomogeneous broadening contribution originating from the sample imperfections and assumed to be frequency independent.

\begin{figure*}[h]
\centering
\includegraphics[width=\textwidth]{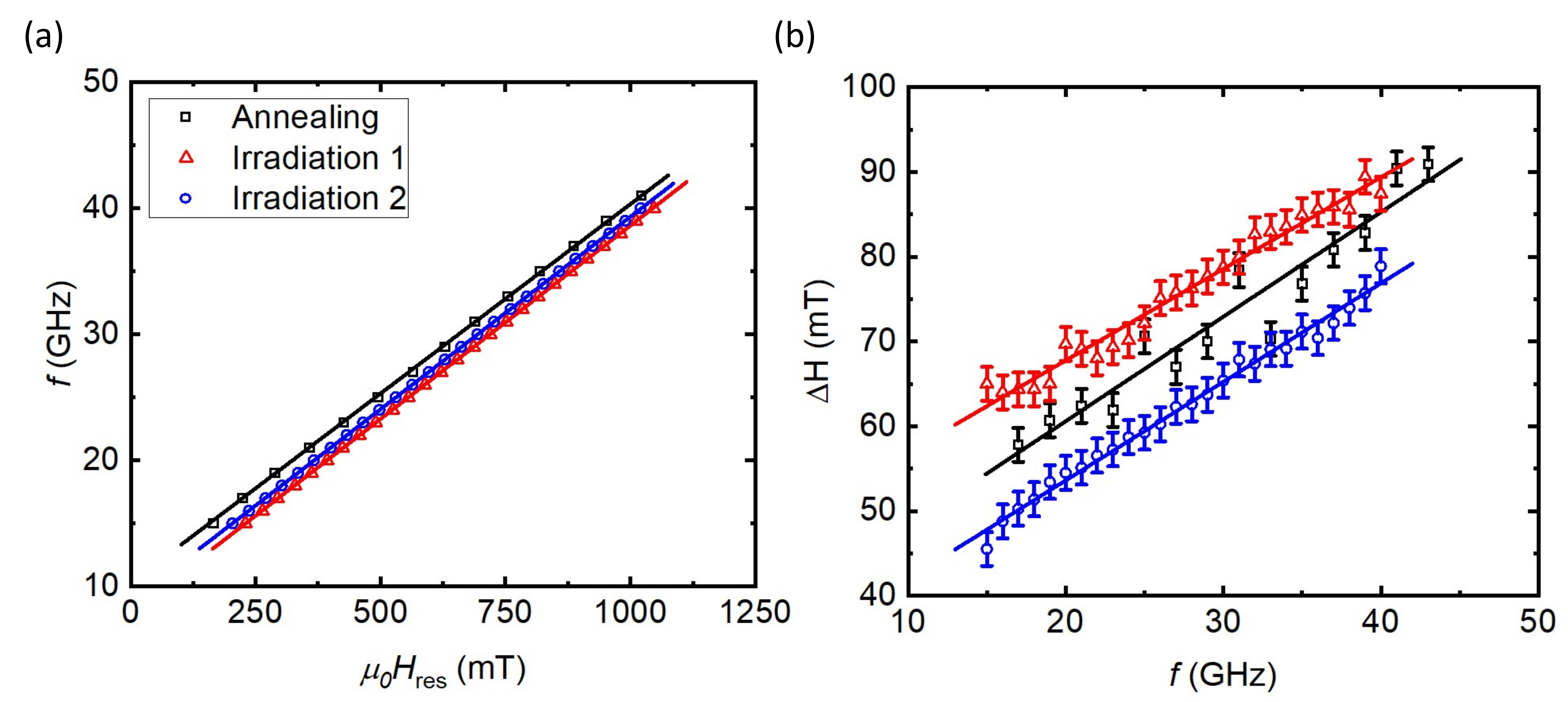}
\caption{(a) Frequency vs the measured perpendicular resonance field for each sample reported in the main paper. (b) Measured linewidth vs frequency. Solid lines in each figure are fits using Eqs. \ref{eq:KittelOOP} and \ref{eq:FMRlinewidth} for (a) and (b) respectively.}
\label{fig:FMR}
\end{figure*}

For the FMR measurements in the main paper, we measured over a frequency range of 15-43 GHz in the out-of-plane direction. The resulting resonant fields and linewidths are shown in figure \ref{fig:FMR}. The measured resonant fields (Fig. \ref{fig:FMR}a) are fitted with Eq. \ref{eq:KittelOOP} to extract $\gamma$ and $H_{K_\mathrm{eff}}$. The linewidths (Fig. \ref{fig:FMR}b) are fitted with Eq. \ref{eq:FMRlinewidth} to extract $\alpha$ and $\Delta H_\mathrm{0}$.

\subsection{Brillouin Light Scattering (BLS) measurements}\label{sec:BLS}
BLS is a powerful technique that can be used to obtain information about material properties like the damping, DMI and homogeneity of the magnetic layer. In this section, we describe in more detail how we used BLS to characterize our samples. 

\begin{figure*}[ht]
\centering
\includegraphics[width=.8\textwidth]{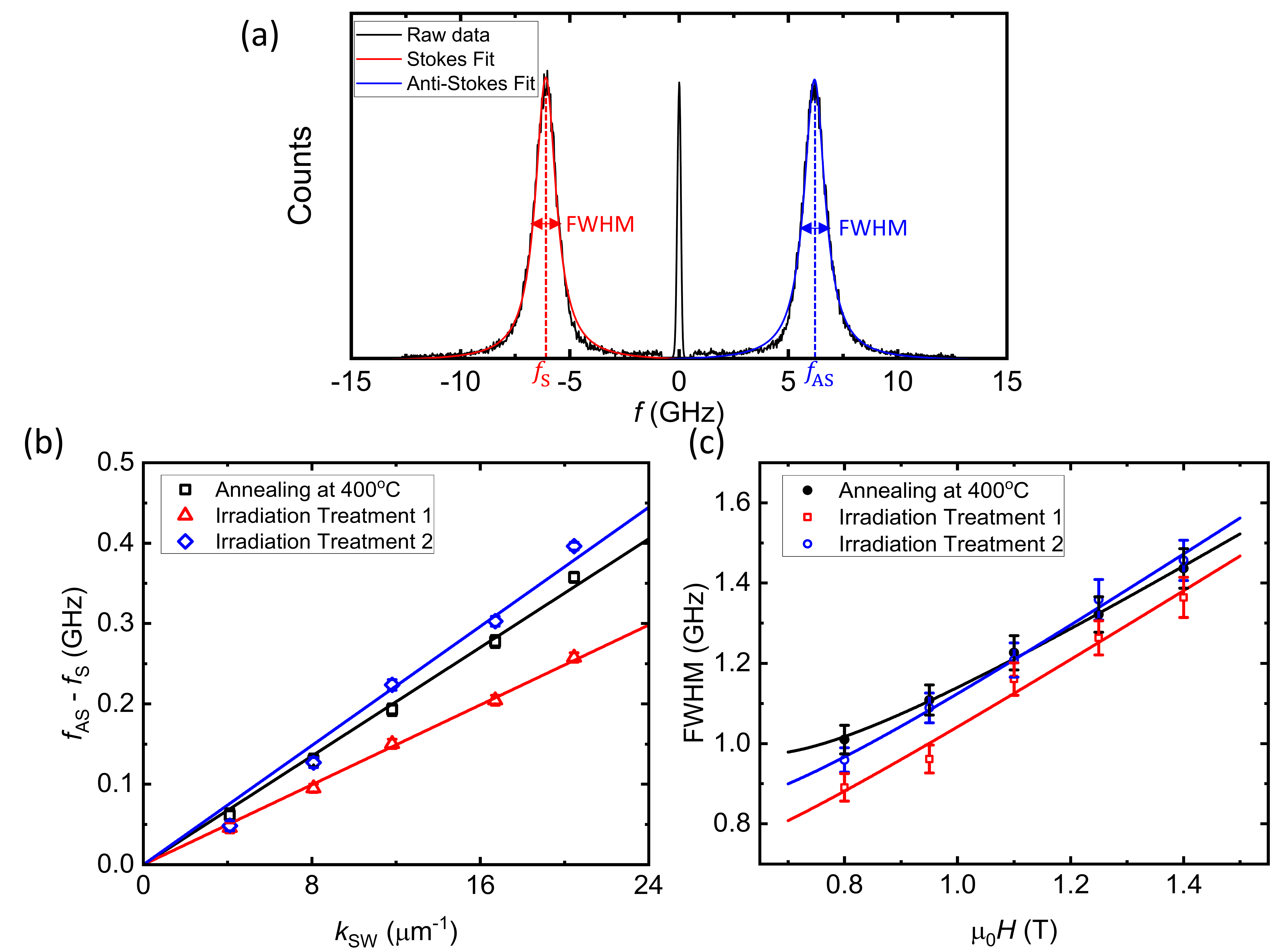}
\caption{(a) Example of a spectrum obtained from BLS measurements. The raw data denoting counts of photons with a specific wave vector is shown in black, and the Lorentzian fits of the Stokes and anti-Stokes lines are shown in red and blue respectively. The central peak has been attenuated so to be on the same height as the Stokes and anti-Stokes peaks. The frequencies and linewidths extracted from the fits are shown for reference purposes. (b) BLS frequency shift as a function of wave vector $k_\mathrm{sw}$ for the three samples presented in the main paper. Solid lines are linear fits using Eq. \ref{eq:BLSDMI}. (c) FWHM of the Lorentzian fits as a function of the applied magnetic field. Solid lines are fits with Eq. \ref{eq:BLSlinewidth}.}
\label{fig:BLS}
\end{figure*}

We applied a magnetic field of $H > H_{K_\mathrm{eff}}$ propagating along the in-plane direction, perpendicular to the direction of the incidence light with $\lambda =$ 532 nm, which allows us to probe spin-waves in the in-plane direction, perpendicular to the applied field. This is the so-called Damon-Eshbach geometry. Furthermore, we obtained spectra for each individual incidence angle after counting photons for up to 15 hours for the largest incidence angles. $f_\mathrm{S}$ and $f_\mathrm{AS}$ are then determined by fitting the peaks with a Lorentzian. (e.g. Fig. \ref{fig:BLS}a).

In systems with DMI, the Stokes and anti-Stokes processes do not have the same energy. DMI favors positive or negative wave-vector spin waves, depending on its sign \cite{moon_spin-wave_2013, Cortes_Ortuno_Influence_2013}. Hence, there is a frequency shift, which depends on the strength of the DMI, and is given by

\begin{align}
f_\mathrm{DMI} = \frac{\gamma}{\pi M_\mathrm{s}} D k_\mathrm{sw},
\label{eq:BLSDMI}
\end{align}

where $D$ is the DMI constant, so that the final frequency of the scattered photon in systems with DMI becomes

\begin{align}
f = f_\mathrm{0} \pm f_\mathrm{DMI},
\end{align}

where $f_\mathrm{0}$ is the frequency of the spin waves in systems without DMI, given by the expression reported in ref. \cite{belmeguenai_interfacial_2015}.

To extract the DMI strength, we only need to measure the frequency difference between the Stokes and Anti-Stokes peaks $\Delta f = f_\mathrm{AS} - f_\mathrm{S} = 2 f_\mathrm{DMI}$. The spin-wave vector that we probe can be varied by changing the incidence angle $\theta_\mathrm{inc}$ of the laser, since $k_\mathrm{sw} = 4\pi\sin\left(\theta_\mathrm{inc}\right)/\lambda$. By measuring $\Delta f$ as a function of $k_\mathrm{sw}$ we can extract the DMI strength from the slope.

Figure \ref{fig:BLS}b shows the frequency shift $\Delta f$ as a function of the spin-wave vector for the samples studied in the main paper. Note that we have plotted $f_\mathrm{AS} - f_\mathrm{S}$, due to the left-handed chirality of our system. This makes sure that the sign of the DMI is correct \cite{belmeguenai_interfacial_2015}. The solid lines are fits with Eq. \ref{eq:BLSDMI}, which are forced through 0 to adhere to Eq. \ref{eq:BLSDMI}.

Similarly to FMR, we can use the width of the peaks in the BLS spectrum to obtain more information about our sample. Specifically, we can obtain information about the Gilbert damping constant $\alpha$ and about the broadening due to sample disorder. For $H > H_{K_\mathrm{eff}}$, the linewidth, defined as the full width at half maximum (FWHM) of the peaks in the spectra (see also Figure \ref{fig:BLS}a), is given by \cite{capua_determination_2015}:

\begin{align}
\mathrm{FWHM} = \frac{\alpha\gamma_0}{2\pi}\left(2H - H_{K_\mathrm{eff}}\right) + \frac{\gamma_0 H}{4\pi\sqrt{H^2 - H H_{K_\mathrm{eff}}}}\Delta H_{K_\mathrm{eff}},
\label{eq:BLSlinewidth}
\end{align}

where the first term represents the contribution from the damping, and the second term represents the contribution of the disorder in the materials, modeled through $\Delta H_{K_\mathrm{eff}}$, which is the distribution of effective anisotropy fields.

Figure \ref{fig:BLS}c shows the linewidth as a function of the applied field, where we already subtracted the contribution of the spectrometer (0.38 GHz, equal to the width of the central peak that arises from the finesse of the spectrometer cavity, see Fig. \ref{fig:BLS}a). The solid lines are fits with Eq. \ref{eq:BLSlinewidth}, where $\alpha$ and $\Delta H_{K_\mathrm{eff}}$ are used as free fitting parameters. Dividing $\Delta H_{K_\mathrm{eff}}$ by the effective anisotropy field value gives us an estimate for the percentage of disorder in the samples. The values are reported in Table II of the main paper and in Table \ref{tab:linewidths}.

\section{Saturation magnetization and microscopic pinning parameters}\label{sec:satmag}
The samples in the main paper have very similar micromagnetic properties, except for the saturation magnetization. Here we show that the results from the main paper do not depend significantly on the value of the saturation magnetization.

\begin{figure}[t]
\centering
\includegraphics[width=0.5\textwidth]{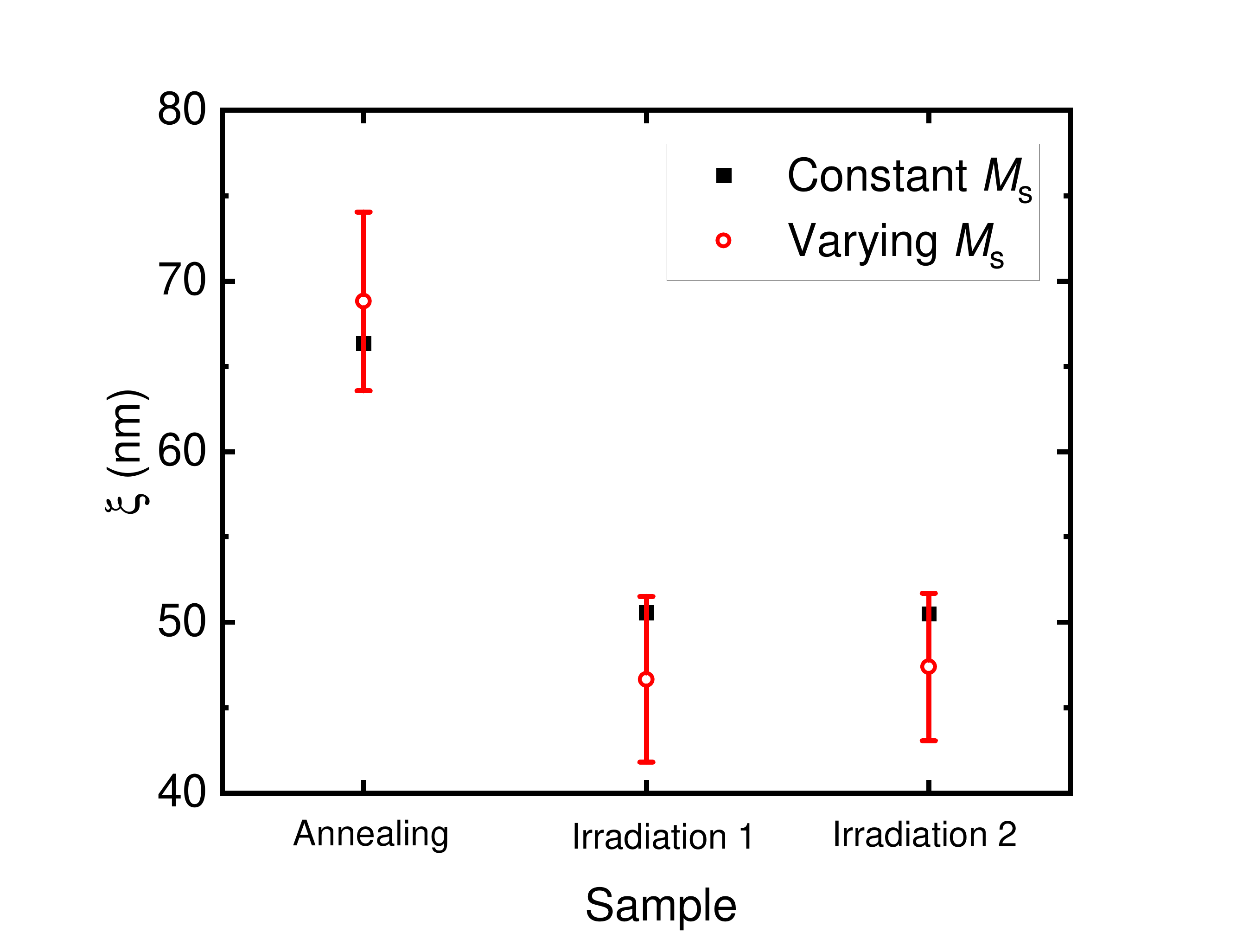}
\caption{Microscopic pinning ranges in the three investigated samples from two different analyses. The analysis with constant $M_\mathrm{s}$ is shown in black, and the analysis with varying $M_\mathrm{s}$ from the main paper is shown in red.}
\label{fig:satmagxi}
\end{figure}

We perform the same analysis, but now with constant $M_\mathrm{s} = 1.59$ MA/m. We also assumed $A = 10$ pJ/m, but the rest of the parameters are the same. Figure \ref{fig:satmagxi} shows the resulting values for the microscopic pinning length $\xi$ from the two analysis. The analysis with constant $M_\mathrm{s}$ is shown in black, while the results with varying $M_\mathrm{s}$ are shown in red.

Figure \ref{fig:satmagxi} clearly shows that the resulting values for $\xi$ are very similar, and not strongly dependent on the value of $M_\mathrm{s}$, with a similar trend of reduced $\xi$ for both analyses. The effect is similar for $f_\mathrm{pin}$ and $\frac{1}{\sqrt{n'}}$, where the trend for both analyses is the same. Furthermore, the extracted values for the creep energy barrier $k_\mathrm{B}T_\mathrm{d}$ come from the domain wall motion curve, and are thus independent of the value of $M_\mathrm{s}$. From these results we can confirm that the conclusions regarding the pinning energy landscape from the main paper are not affected.

\section{Kerr Microscopy procedure}
\begin{figure*}[ht]
\centering
\includegraphics[width=\textwidth]{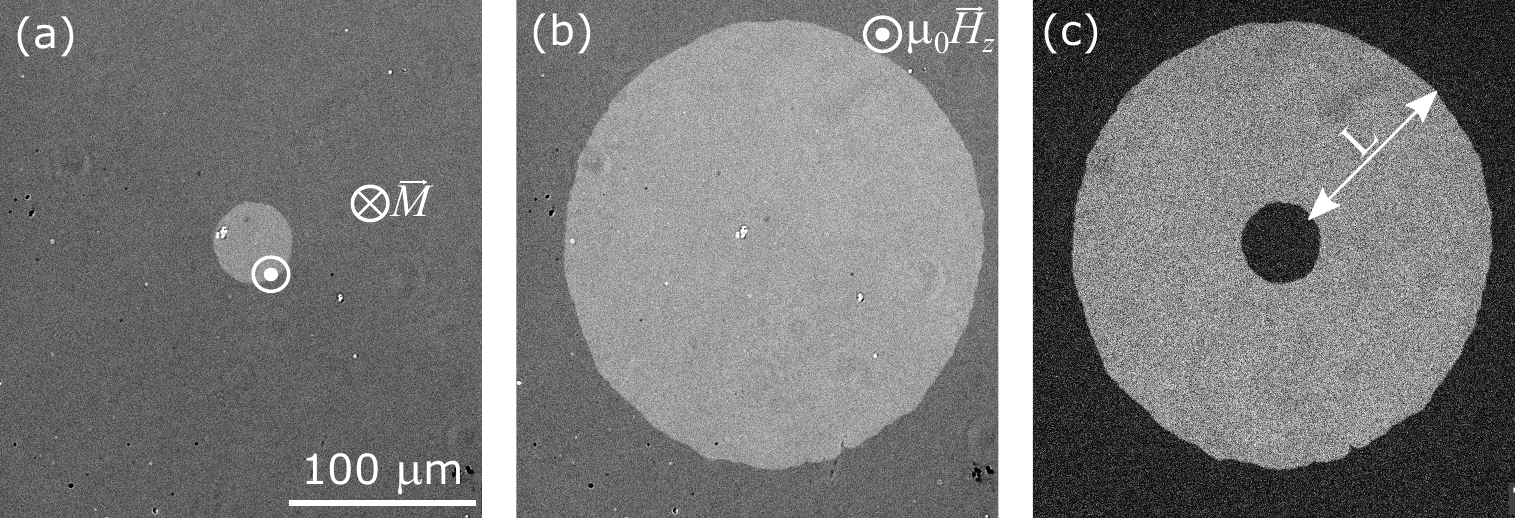}
\caption{Example measurement on the determination of the domain wall velocity. (a) A single domain is nucleated (shown in light grey), where the magnetization around the domain (dark grey) is in the opposite direction. (b) The same domain after a magnetic field pulse has been applied in the $z$-direction. (c) Differential image (b) - (a). The white area now denotes the expansion of the domain during the magnetic field pulse. The distance $L$ denotes the displacement length of the domain wall. Figure inspired by ref \cite{caballero_excess_2017}.}
\label{fig:KerrMicroscopyProcedure}
\end{figure*}

For the measurements in the paper we use a home made Kerr microscope and coils with a short rise time of roughly 100 ns and a diameter of a few mm. The procedure for measuring the domain wall motion by Kerr microscopy is shown in Figure \ref{fig:KerrMicroscopyProcedure}. We first saturate the sample with a strong and long out-of-plane magnetic field pulse. We then nucleate the initial domain with a strong and short pulse in the opposite direction, resulting in the state shown in Figure \ref{fig:KerrMicroscopyProcedure}a, where the initial domain is shown in light grey. This is followed by a magnetic field pulse of up to 45 mT with lengths down to 500 ns in the same direction as the initial domain, causing it to grow. Another image is taken after this, resulting in the state shown in Figure \ref{fig:KerrMicroscopyProcedure}b. Figure \ref{fig:KerrMicroscopyProcedure}a is then subtracted from Figure \ref{fig:KerrMicroscopyProcedure}b to obtain the differential image shown in Figure \ref{fig:KerrMicroscopyProcedure}c. Note that the light grey area now represents the area over which the domain wall has displaced itself. By measuring the distance denoted by $L$ in Figure \ref{fig:KerrMicroscopyProcedure}c, we can extract the domain wall velocity $v_\mathrm{DW} = L/t_\mathrm{p}$, where $t_\mathrm{p}$ is the duration of the pulse. To obtain good statistics, at least 10 different measurements of $L$ are taken in all directions.

\section{BLS vs FMR inhomogeneous broadening}
In the main text, we mention that the values for the inhomogeneous broadening obtained from BLS and FMR do not agree. In this section, we will elaborate and discuss potential causes. Table \ref{tab:linewidths} contains the results from the FMR measurements and analysis using Eqs. \ref{eq:KittelOOP} and \ref{eq:FMRlinewidth}, as well as the results of the BLS measurements and analysis using Eq. \ref{eq:BLSlinewidth} and the relation for the frequency reported in ref \cite{belmeguenai_interfacial_2015}.

\begin{table}[t]
\centering
\caption{Parameters used for, and extracted from, the linewidth measurements using FMR and BLS. The table reports values for the gyromagnetic ratio $\gamma / 2\pi$ and
the effective anisotropy field $H_{K_\mathrm{eff}}$ extracted from FMR and BLS. The table also reports the results of the analysis of the FMR linewidth measurements using Eq. \ref{eq:FMRlinewidth}, namely $\alpha^{\mathrm{FMR}}$ and $\Delta H_\mathrm{0}$, and $\Delta H^{\mathrm{FMR}}_{K_\mathrm{eff}} / H^{\mathrm{FMR}}_{K_\mathrm{eff}}$, as well as the results from analysis of the BLS linewidth measurements using Eq. \ref{eq:BLSlinewidth}, namely $\alpha^{\mathrm{BLS}}$, $\mu_\mathrm{0}\Delta H^{\mathrm{BLS}}_{K\mathrm{eff}}$ and $\Delta H^{\mathrm{BLS}}_{K_\mathrm{eff}} / H^{\mathrm{BLS}}_{K_\mathrm{eff}}$.}
\begin{tabular}{c c c c}
\hline \hline
 & Annealing & Irradiation 1 & Irradiation 2 \\
\hline
$\gamma^{\mathrm{FMR}} / 2\pi$ (GHz/T) & 30.1 $\pm$ 0.1 & 30.67 $\pm$ 0.02 & 30.55 $\pm$ 0.02\\
$\mu_\mathrm{0}H^{\mathrm{FMR}}_{K_\mathrm{eff}}$ (mT) & 342 $\pm$ 6 & 259 $\pm$ 1 & 287 $\pm$ 1\\
$\alpha^{\mathrm{FMR}}$ (10$^{-3}$) & 19 $\pm$ 1 & 17 $\pm$ 1 & 18 $\pm$ 1\\
$\mu_\mathrm{0}\Delta H_\mathrm{0}$ (mT) & 36 $\pm$ 3 & 46 $\pm$ 1 & 30 $\pm$ 1 \\
\\
$\gamma^{\mathrm{BLS}} / 2\pi$ (GHz/T) & 29.1 $\pm$ 0.6 & 29.3 $\pm$ 0.7 & 29.3 $\pm$ 0.6\\
$\mu_\mathrm{0}H^{\mathrm{BLS}}_{K_\mathrm{eff}}$ (mT) & 426 $\pm$ 27 & 313 $\pm$ 38 & 367 $\pm$ 30\\
$\alpha^{\mathrm{BLS}}$ (10$^{-3}$) & 15 $\pm$ 1 & 15 $\pm$ 1 & 16 $\pm$ 1\\
$\mu_\mathrm{0}\Delta H^{\mathrm{BLS}}_{K\mathrm{eff}}$ (mT) & 24 $\pm$ 1 & 16 $\pm$ 3 & 20 $\pm$ 1\\
$\Delta H^{\mathrm{BLS}}_{K_\mathrm{eff}} / H^{\mathrm{BLS}}_{K_\mathrm{eff}}$ (\%) & 5.6 $\pm$ 0.6 & 5.3 $\pm$ 1.7 & 5.3 $\pm$ 0.8\\
\hline
\end{tabular}
\label{tab:linewidths}
\end{table}

Regarding the reduced gyromagnetic factor $\gamma / 2\pi$, we observe that the FMR values are very close with published measurements on nominally identical samples, for example 30.32 $\pm$ 0.04 GHz/T for the annealed sample \cite{devolder_damping_2013}. The lower factor in the in-plane geometry measured by BLS is also consistent with the orbital contribution to the gyromagnetic ratio expected for samplses with perpendicular easy axis \cite{shaw_precise_2013}.

Bigger differences are observed for the effective anisotropy field. They can be ascribed to the presence of a higher order contribution to the PMA, which is not detected in the perpendicular geometry but modifies the in-plane resonance frequency. For the samples considered here, the sign of the second order PMA appears to be positive, with values of the corresponding anisotropy field of 84, 54 and 80 mT for the annealed sample and the samples with irradiation treatment 1 and irradiation treatment 2, respectively. Another possible cause of the discrepancy between OOP and IP FMR could be the presence of a magnetic anisotropy within the sample plane, but such a term is not expected as the samples were grown under rotation within the sample plane. A last matter of concern is that the BLS measurements were performed at a finite wavevector $k_\mathrm{SW} = 11.81$ $\mu$m$^{-1}$. Comparing with the Kittel equation at $k_\mathrm{SW} = 0$, one sees that a finite $k$ amounts to replacing, in the ultra thin limit, the applied field $H$ by $H + \frac{1}{2}ktM_\mathrm{s} + \frac{2A}{\mu_\mathrm{0}M_\mathrm{s}}k^{2}$, and the anisotropy field $H_\mathrm{K}$ by $H_\mathrm{K} + ktM_\mathrm{s}$. Using the magnetization values reported in Table 1 of the main text, one obtains for the linear term in $H$ (times $\mu_\mathrm{0}$) values of 9, 11 and 12 mT, whereas for the quadratic term the values are all 3 mT. Therefore, the difference of anisotropy fields between FMR and BLS is real, and can be explained by the second order PMA constant.

Regarding the linewidth measurements, we note that FMR and BLS result in very similar values for the damping parameter, with $\alpha^{\mathrm{FMR}}$ maybe slightly larger than $\alpha^{\mathrm{BLS}}$. However, larger differences are present when looking at the inhomogeneous broadening $\mu_\mathrm{0}\Delta H_\mathrm{0}$ and dispersion of the anisotropy field $\mu_\mathrm{0}\Delta H^{\mathrm{BLS}}_{K\mathrm{eff}}$. The values obtained from both techniques do not agree, with $\mu_\mathrm{0}\Delta H_\mathrm{0}$ larger than $\mu_\mathrm{0}\Delta H^{\mathrm{BLS}}_{K\mathrm{eff}}$. Note that, in the perpendicular geometry where FMR was performed, the additional relaxation channel called two-magnon scattering \cite{mills_spin_2003} is absent, so that $\mu_\mathrm{0}\Delta H_\mathrm{0}$ should be the true inhomogeneous broadening of the resonance line. For BLS, even if it is performed in the in-plane geometry, where the two-magnon scattering channel is active, no associated increase of the linewidth is expected. Indeed, magnons are not pumped at $k = 0$ in BLS, but probed at a specific $k$. From this analysis therefore, we expect that $\mu_\mathrm{0}\Delta H_{K\mathrm{eff}}$, evaluated by removing the non-linearity of in-plane resonance, should be equal to $\mu_\mathrm{0}\Delta H_\mathrm{0}$. In fact, we would even expect that $\mu_\mathrm{0}\Delta H_{K\mathrm{eff}} > \mu_\mathrm{0}\Delta H_\mathrm{0}$ as in the in-plane configuration the second PMA constant is active, whose dispersion adds to that of $M_\mathrm{s}$ and $K_\mathrm{u}$. The experimental results, even when taking into account the larger error bars on $\mu_\mathrm{0}\Delta H_{K\mathrm{eff}}$ due to the small number of data points, are in stark opposition to these expectations, especially for the sample with Irradiation treatment 1. This definitely calls for further studies. Despite these discrepancies, the values for $\Delta H^{\mathrm{BLS}}_{K_\mathrm{eff}} / H^{\mathrm{BLS}}_{K_\mathrm{eff}}$ obtained from BLS agree with previous work on similar samples \cite{herrera_diez_enhancement_2019}, and hence we use them for the analysis in the main paper where anisotropy variations come into play (e.g. in Section IIIc).

\section{Micromagnetic Simulations in the flow regime}
To shed light on the variation of flow velocity in the three samples in the main paper, we performed an extensive set of micromagnetic simulations using Mumax3 \cite{vansteenkiste_design_2014, leliaert_current-driven_2014, mulkers_effects_2017}. In the same vein as ref \cite{voto_effects_2016}, we investigated domain wall velocity in the precessional regime in disordered samples. We will use the results of these simulations to argue that the variation of the flow velocity in the three samples in the main paper cannot simply be explained by variation of magnetic parameters, and that instead the disorder has a role to play.

First, we study the effect of the disorder parameters on the flow velocity, where we focus specifically on the grain size and distribution of magnetic parameters. The system under study is the annealed sample from the main paper, and thus we have used the magnetic parameters reported in Table 1 of the main paper ($M_\mathrm{s} =$ 1.25 MA/m, $K_\mathrm{u} =$ 1.20 MJ/m$^{3}$, $\alpha =$ 0.019, $D =$ 0.18 mJ/m$^{2}$, and $A =$ 15 pJ/m). The computational window is $1 \times 2 \mu\mathrm{m}^{2}$, with $2 \times 2 \mathrm{nm}^{2}$ cells and a thickness of 1 nm. Grains are introduced using a Voronoi tessellation \cite{leliaert_current-driven_2014}, and the grains are subsequently given a distribution of anisotropy values around their nominal value $K_\mathrm{u}$ with a distribution $\sigma$, and a deviation of anisotropy axis around the $z$-direction with the same distribution $\sigma$. Variations of other parameters are not considered, and the simulations are performed without the influence of temperature. 

\begin{figure}[ht]
\centering
\includegraphics[width=\textwidth]{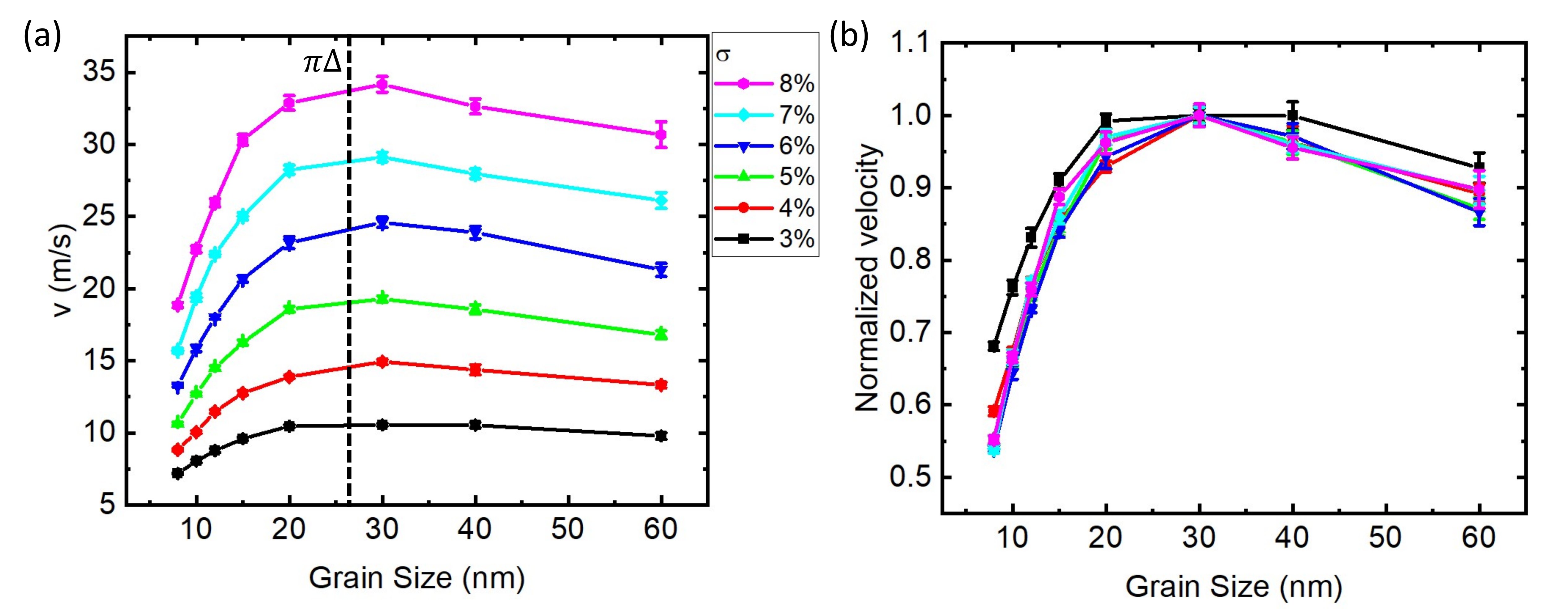}
\caption{(a) Micromagnetic simulations of the domain wall velocity at 70 mT as a function of the average Voronoi grain size for differing anisotropy distribution values $\sigma$. The dashed line represents the Bloch domain wall width $\pi \Delta = \pi \sqrt{A/K_\mathrm{eff}}$ = 26 nm. The velocity is averaged over 10 different grain distributions, and the error bars denote the standard deviation in the average velocity over these grain distributions. (b) Data of (a) normalized at a grain size of 30 nm.}
\label{fig:mumaxgrains}
\end{figure}

We investigate the velocity at an applied perpendicular field $B_z =$ 70 mT, which is well above the depinning transition, but below the threshold for nucleation in our sample. Furthermore, the domain wall motion is in the precessional regime. We vary the average Voronoi grain size between 8 and 60 nm, for values of $\sigma$ ranging between 3 and 8\%. Each simulation is averaged over 10 different realizations of the grain distribution and anisotropy distribution. The results of this study are shown in Figure \ref{fig:mumaxgrains}.

Figure \ref{fig:mumaxgrains}a shows that for all values of $\sigma$, the velocity increases sharply with increasing grain size up to roughly the value of the Bloch domain wall width $\pi \Delta = \pi \sqrt{A/K_\mathrm{eff}}$ = 26 nm, after which it slightly decreases. This is in good agreement with previous simulations \cite{voto_effects_2016, kim_current-driven_2017}. Furthermore, figure \ref{fig:mumaxgrains} shows that the DW velocity in the flow regime is strongly impacted by the strength of the dispersion $\sigma$, with larger dispersions resulting in faster DW motion. This is in agreement with previous simulations \cite{min_effects_2010, leliaert_current-driven_2014} and can be explained with an extra extrinsic component to the effective damping. The system dissipates energy through the emission of a spin wave at local DW depinning events, increasing the DW velocity \cite{voto_effects_2016}. The curves in Figure \ref{fig:mumaxgrains}a have a similar shape. Figure \ref{fig:mumaxgrains}b shows the same data, but now normalized at 30 nm, revealing that the grain size has almost exactly the same effect no matter the strength of the anisotropy distribution.

\bibliography{bibliographysupplementalmaterial}